\documentstyle[preprint,epsfig,aps]{revtex}

\begin{document}

\centerline{\bf Cherenkov radio pulses from EeV neutrino interactions: the LPM effect} 


\centerline{J. Alvarez-Mu\~niz and E. Zas.}

\centerline{\it Departamento de F\'\i sica de Part\'\i culas, Universidad de
Santiago}
\centerline{\it E-15706 Santiago de Compostela, Spain}

\begin{abstract}
We study the implications of the LPM effect for the Cherenkov radiation of EeV electromagnetic showers in the coherent radiowave regime for ice. We show 
that for showers above 100~PeV the electric field scales with shower energy but 
has a markedly narrower angular distribution than for lower energy showers. We 
give an electric field frequency and angular spectrum parameterization valid 
for showers having energy up to the EeV regime and discuss the 
implications for neutrino detectors based on arrays of radio antennas. Implications of the LPM effect 
for under water neutrino detectors in project are also briefly addressed.
\end{abstract}

{\bf PACS number(s):} 96.40.Pq 29.40.-n; 96.40.Kk; 96.40.Tv 

{\bf Keywords:} Cherenkov radiation, LPM effect, 
Electromagnetic showers, Neutrino detection.

\section{Introduction}

Neutrino detection is one of the most exciting challenges in particle astrophysics. 
Low energy neutrinos from astrophysical environments have already been detected from 
the sun and from the supernova explosion SN 1987A, opening up a new window to 
the Universe. High energy neutrinos have to be produced in the interactions of cosmic rays 
with both matter or radiation and are also very likely to be produced by whatever mechanism 
that accelerates protons or nuclei to cosmic ray energies. 
Gamma Ray Bursts, Active Galactic Nuclei
and topological defect annihilation are sites where hadrons may be accelerated  
to energies above $10^{19}~$eV, but this is still 
a matter of speculation and intense research \cite{Auger}.
Many of the proposed models for acceleration in these sites also predict distinct 
neutrino fluxes which extend above the EeV range \cite{Biermann}. 
In any case cosmic rays above few $10^{19}~$eV (believed to be extragalactic) 
must produce high energy neutrinos in their interactions with the cosmic microwave 
background, whatever their origin, at flux levels that are   
challenging low \cite{Berezinskii}.  
Neutrinos of energies as high 
as observed in cosmic rays can be expected at flux levels which will be uncertain until 
the cosmic ray acceleration mechanism is identified and well 
understood \cite{zasmoriond}. Their detection would
provide extremely relevant information for the establishment of the origin of the highest
energy cosmic rays. 

There are several ``conventional '' projects to instrument large volumes of water or ice 
with photodetectors \cite{physrep} (two in operation \cite{Amanda}). They can detect
Cherenkov light   
exploiting the long range of upcoming muons produced in charged current 
neutrino interactions.  
The search of EeV neutrinos with these detectors implies the detection of showers. 
As the earth is opaque to EeV neutrinos only vertical downgoing to 
horizontal neutrinos can be observed. The atmospheric muon background must be eliminated 
to identify neutrino induced muons, looking for energetic showers developed along the muon 
track. Alternatively the Cherenkov light can be also detected from the showers 
directly produced in the neutrino interactions. This is the Cherenkov light  
emitted incoherently by shower particles in the optical
band.  

The search for coherent radio pulses generated in ice by the charge imbalance in 
neutrino induced electromagnetic showers provides an interesting alternative, 
\cite{zele} known since the 1960's \cite{aska}, 
that may turn out to be more cost effective at EeV 
energies \cite{price}. The Cherenkov radiation of shower particles 
is coherent when the wavelength of the radiation is greater than the physical dimensions 
in the shower. The radiated electric field becomes proportional to the square of the excess charge that develops in the shower and 
the power in the radio emission scales  
with the square of the shower energy \cite{aska}. Simulations of radio pulses from these showers have 
been made up to 10~PeV \cite{Zas} which indicate that only neutrinos above few PeV 
can be detected by a single antenna at distances in the km range. In spite of 
experimental difficulties still unexplored\cite{jelley}, 
the technique is very attractive to detect 
neutrinos of energy in the EeV range and above and there are efforts under way to test 
its viability \cite{RICE}. The mechanism also suggests that it may be possible to detect 
neutrino radio pulses produced under the moon surface using radiotelescopes on earth 
\cite{jaime}. 

The development of electromagnetic showers in dense media is of utmost importance for 
the detection of EeV neutrinos from both conventional underground and radio pulse detectors. 
The highest energy showers generated along a muon track are mostly due to bremsstrahlung 
because it has the hardest cross section and hence they will be electromagnetic. 
In charged current electron neutrino 
interactions the electron at the lepton vertex carries on average most of the energy 
\cite{reno} and the induced shower will also have a dominant electromagnetic character.  
Electromagnetic EeV showers in water or ice are dramatically affected by the Landau 
Pomeranchuck Migdal effect (LPM) \cite{LPM,migd} which is a manifestation of the collective 
electric field of matter at large distances \cite{konishi,stanvan,Misaki}. In this work we 
consider the development of EeV electromagnetic showers in water and the implications for 
Cherenkov radiation both in the incoherent (optical) and coherent (radio pulses) regions 
which are respectively relevant for the conventional underground muon detectors and the 
radio technique for EeV neutrino detection. Hadronic showers will be less affected by the
LPM effect and will be addressed elsewhere. 

\section{The LPM effect}

The LPM effect is due to the peculiar bremsstrahlung and pair production kinematics for 
which the average interaction distance is proportional to interaction energy \cite{Nishimura}. 
When the energy of the incoming photon or electron particle gets sufficiently high, the 
interaction distance becomes comparable to the 
interatomic spacing and collective atomic and molecular effects affect the static electric 
field responsible for the interaction. This naturally introduces an energy scale above 
which these effects become significative, $E_{LPM}$ \cite{stanvan} which is highly dependent 
on the density of the medium. The result is a dramatic reduction in both total cross 
sections with lab energy ($E$) which drop like $E^{-0.5}$ above $\sim E_{LPM}$. 
It also suppresses the central part of the differential cross section for pair production 
(where the electron and positron carry similar fractions of the incoming photon energy) 
and cuts the cross 
section off for bremsstrahlung of very low energy photons \cite{konishi,stanvan}. 
 
As a result the mean free path of an EeV photon or electron is 
considerably larger than for PeV energies and the shower develops 
further away from injection.
This is irrelevant for neutrino detection 
because the low cross sections involved make most points around the detector equally 
probable for a neutrino interaction. Secondary interactions will also be separated by depth 
intervals that are considerably longer than the radiation length as long as they are 
induced by particles of energy above $E_{LPM}$. This elongates the shower's depth 
distribution with respect to lower energy showers. At EeV energies there is also a 
tendency for photons to pair 
produce electron-positron pairs with a markedly leading particle (more likely to have 
energy above $E_{LPM}$), which contributes to elongate the shower even further. The resulting 
showers are markedly different from standard showers obtained in shower theory 
\cite{Nishimura}. 
As the interaction length of high energy particles is larger than in standard showers,  
showers affected by LPM 
fluctuate more than conventional ones; this can compromise the possibility of shower 
energy reconstruction.  

\section{Shower simulations}

The brute force simulation of radio pulses from EeV showers involves 
following particles through at 
least twelve orders of magnitude in energy because half of 
the tracklength in an electromagnetic shower in water is due to particles below $3~$MeV, 
which produce interference effects that affect the overall angular and frequency distribution 
of the pulse \cite{Zas}. We have developed a fast 
3D Monte Carlo for electromagnetic showers in ice to run at the highest energies based on 
the one described in \cite{Zas}. 
The Monte Carlo is combined with shower parameterizations to achieve reasonable 
computing times avoiding thinning, because of possible biasing 
of the results due to interference effects. 
We follow in detail particles of energy above a   
threshold $E_{th}$, which is always chosen to be well below $E_{LPM}$, typically 100~TeV 
to run at the highest energies. 
We use the Greisen's parameterization for the subthreshold 
showers to obtain the particle 
depth distributions. Lateral distributions and tracklengths  
are calculated using the NKG parameterization\cite{NKG} 
and the tracklength results obtained in \cite{Zas}. 

Intermediate shower results 
have been checked for consistency and compared to full simulations when the primary 
energy is not impractically large. The full simulations are consistent with 
simulations using $E_{th}$ up to energies below $E_{LPM}$, which constitutes a test of the 
parameterizations used. Results are not significantly affected by changes in $E_{th}$. 
Moreover they are in agreement with other work on LPM showers 
\cite{stanvan,Misaki}.

The main differences with ordinary showers can be summarized as follows: 

\hskip 1cm
1) As discussed in section II, showers of energy above $E_{LPM}$  
display an elongated depth distribution.
 
\hskip 1cm
2) The total tracklength and the difference of electron and positron tracklengths 
projected onto the shower axis (weighted projected tracklength) are very little 
affected by the LPM effect, that is they continue to scale with primary energy to 
very high accuracy. This is not surprising since only high energy particles suffer 
LPM effect and most of the tracklength is due to low energy particles in agreement with  
earlier discussions \cite{ralpm,Misaki}.  

\hskip 1cm
3) The depth distribution around shower maximum becomes considerably longer than the 
Greisen parameterization but the maximum number of particles $N_{max}$ becomes lower. 
The integral in depth of the (excess) number of particles can be related to the (weighted) 
tracklength. The area under the depth distribution curve remains proportional to shower 
energy to a very good approximation. We conveniently introduce {\sl shower length} as 
depth over which a shower has more particles than a given fraction of $N_{max}$. The 
results for the average shower length are displayed in Fig.~1. Significant deviations  
from Greisen showers are observed when the energy exceeds $\sim 10~E_{LPM}$. 
Beyond this point the rise in length with shower energy $E$ can be approximately parameterized as $E^{1/3}$. 

\hskip 1cm
4) The showers display a multi-peak longitudinal structure which    
is due to the change in the 
LPM differential pair production cross section with energy, as well as to the total 
cross section decrease for both pair production and bremsstrahlung \cite{konishi}. 

\hskip 1cm
5) The showers show a lateral distribution which is consistent with the NKG 
parameterization for values of the age parameter $s$ close to $s=1$. 

\hskip 1cm
6) Fluctuations are in general a great deal larger than in regular showers and they 
get enhanced with energy. Shower fluctuations to $N_{max}$ larger than the average 
value correlate with shorter cascades. These are strong correlations that can 
be related to point 3). 
Fig.~2 displays the distributions of shower length as defined in 3) for a sample of 200 
showers illustrating the enhancement in the large length tail as the shower energy rises.  

\vskip 0.3cm

\section{Cherenkov radiation}

The implications for optical Cherenkov emission (incoherent regime) are 
straightforward. 
The total light output is proportional to the tracklength and continues to scale 
with shower energy also for showers where the LPM effect is very important. When this behavior is combined with the shower length increase discussed in the previous section, the light output per unit area is expected to rise as $E^{\sim 2/3}$. As the length becomes comparable or even longer than 
the typical attenuation length of light in water or ice, the effective volume for shower 
detection of an underground neutrino telescope is increased accordingly. Reconstruction of 
shower energy will however require that the shower is contained in the detector so that the 
total light output can be sampled. For showers totally or partially outside the instrumented
 volume the fluctuations in $N_{max}$ will make energy reconstructions much more uncertain.  

The consequences for radio detection are more subtle. 
The radio pulse spectrum has a complicated shape which can be interpreted as a 
Fraunhofer diffraction pattern of the charge excess in shower as discussed 
in ref. \cite{Zas}. Basically the electric field spectrum rises linearly 
with frequency until the destructing interference effects start 
to take place inducing a maximum at $\nu_{max}$. For pulses emitted at the Cherenkov 
angle with respect to shower axis, this interference is governed by the lateral 
distribution of the shower, the narrower the shower lateral distribution 
the higher $\nu_{max}$. 
This frequency plays the role of an upper limit of the frequency integral for the total 
energy of the pulse. The angular distribution displays a diffraction pattern around the 
Cherenkov angle with a half width of the central peak of $\lambda / L$ where $L$ is 
shower length (using $\sim 0.7~N_{max}$,  
see Fig.~2). As the shower length rises only logarithmically with 
shower energies below $E_{LPM}$, the angular spread of the pulse practically only depends 
on frequency for such energies. Finally the normalization of the spectrum is determined by the weighted 
projected tracklength: the difference of tracklength of negative and positive charges 
after projection onto the shower axis, because they interfere destructively. 

Several cross checks have been implemented to further confirm these interpretations. 
Firstly the Monte Carlo threshold is raised to energies orders of magnitude above the 
nominal 3 MeV. The resulting showers have shorter tracklengths (in agreement with the 
tracklength-threshold relation provided in Fig.~5 of ref.\cite{Zas}) and much steeper lateral 
distributions but the depth distributions still display the typical multipeaked 
and elongated shapes and the same enhancement associated to LPM showers \cite{konishi}. 
The radiopulses generated have a higher  
$\nu_{max}$ than those from showers with thresholds in the MeV range because of 
the steepness of the lateral distribution. The electric field 
at the Cherenkov angle remains proportional to the weighted tracklength reduction. 
The width of the pulse angular distribution 
around the Cherenkov angle becomes smaller than those obtained in the full simulation of 
lower energy showers because of the enhancement in shower length, this effect is truly due 
to the LPM effect and is observed not to depend on the $E_{th}$. 
 
As an alternative check we have artificially enhanced the LPM effect in the medium, 
lowering $E_{LPM}$. This allows simulations with a  
threshold in the nominal MeV range, for lower energy showers displaying (fictitious) 
LPM characteristics. Fig.~1 also illustrates the rise in shower length for 
energies above $\sim 10~E_{LPM}$ which we have chosen to be $\sim~10$ TeV 
in this case. At the Cherenkov peak the generated 
radio pulse scales with weighted tracklength and the angular width of the diffraction 
peak is reduced in proportion to the increase in shower length. The value of 
$\nu_{max}$ remains at a similar value to a real shower of the same energy (not 
affected by the LPM). 
All the results confirm the diffraction pattern interpretation. Moreover the field 
spectrum at angles close to the Cherenkov angle can be reproduced with a Fourier transform 
of the depth distribution. Results will be presented
elsewhere.

The implications of the LPM effect for relevant shower parameters have been discussed 
in section III and can be used to infer with confidence the pulse characteristics for large 
electromagnetic showers that are strongly affected by the LPM effect.  
In summary the electric field value at the Cherenkov peak for EeV showers continues to 
scale with the shower energy as for ordinary showers below $E_{LPM}$ and the electric 
field can be parameterized as \cite{Zas}:
\begin{equation} 
R \vert \vec E(\omega, R,\theta_C) \vert=
1.1 \times 10^{-7}~{\rm E_0 \over 1~{\rm TeV}}~{\nu \over \nu_0}~ 
{1 \over 1 + 0.4 ({\nu \over \nu_0})^2} \; {\rm V~MHz}^{-1}
\end{equation}  
where $\vec E$ is the electric field, $R$ is the observation distance and $\rm E_0$ 
is the shower energy, with $\nu_0 \simeq 500$~MHz. The angular distribution width at 
the Cherenkov peak is reduced in proportion to $\rm E_0^{1/3}$ 
for energies above $\sim 10~E_{LPM}$:
\begin{equation}
E(\omega,R, \theta)=E(\omega,R, \theta_{C})~
\rm e ^ {- {ln 2} \left[ { \theta -\theta_{C} \over \Delta \theta}  
\right]^2 }  
\,\,\,\,\,\,\, {\rm with} \,\,\,\,\,\,\,
\Delta \theta \simeq\cases{
2.7^{\rm o} ~ {\nu_0 \over \nu} ~{\rm E_0}^{-0.03} \,\,{\rm for}\,\, {\rm E_0}<1~PeV~\cr 
2.7^{\rm o} ~ {\nu_0 \over \nu} ~ 
\left[{E_{LPM} \over 0.14~{\rm E_0} + E_{LPM}} \right]^{0.3}
\,\,{\rm otherwise}\,\,}
\end{equation}
Fluctuations in the angular spread can be easily deduced from fluctuations in shower 
length. 

In the absence of attenuation of the pulses, a simple relation can be obtained 
for the maximum distance from which a shower can be detected. This distance scales with 
shower energy for given antenna parameters \cite{Zas}. The relation can be obtained 
dividing the power in the signal at the Cherenkov angle where 
it is maximum by the noise power. If we characterize an antenna 
reception system by its operating frequency 
$\nu_{op}$, its bandwidth $\Delta \nu$ and its equivalent white noise temperature $T$ 
(in K) and we demand a signal-to-noise ratio given by $s/n$, the following distance to shower 
energy relation can be obtained \cite{Zas}:
\begin{equation}
{\rm E_0}({\rm PeV}) \simeq 5~f~ {\sqrt{s/n~T}}~{1 + 1.6\times10^{-6} 
[\nu_{op}({\rm MHz})]^2 
\over {\sqrt{\Delta \nu ({\rm MHz})}}} R ({\rm km})
\end{equation}
We have assumed that the bandwidth is small compared to the frequency of 
observation and that the effective area of the antenna is $\lambda ^2 / 8$ as for a 
half wave dipole\cite{jelley}. Somewhat larger ranges can be obtained with a "TEM horn" 
\cite{prov} or biconical antennas \cite{Ralston}. 
The factor $f$ is a fractional reduction of the electric field due to observation away from 
the Cherenkov direction, detection within half width corresponds to $f=2$. 

A signal-to-noise ratio of 10:1 in a single antenna operating at 1~GHz with bandwidth 
$\sim 100~$MHz and equivalent noise temperature of $300~$K (consistent with 
measurements \cite{bold}) would detect showers above 70 PeV at distances 
below 1 km. Corresponding energy thresholds for lower frequencies and bandwidths are 
given in table I. At lower frequencies the loss in the relation comes because 
of the reduced bandwidth implied by lower operation frequencies. 

For a power reduction by a factor 
$\eta=e^{2 \alpha R}$ with $\alpha$ the attenuation coefficient, Eq.~(3) has an extra  
factor of $\sqrt \eta$. Measurements of the attenuation distance in ice cores 
\cite{Icebook} are tabulated in table II 
as the distance at which 
the power spectrum is halved for several frequencies. This attenuation length increases 
as the temperature drops. Final results however should be corrected when measurements are 
performed "in situ". Absorption in the ice prevents 
detection of pulses from EeV showers at distances where otherwise the signal-to-noise 
ratio requirement would be well satisfied.  
In any case it 
is reasonable to assume that the range of one of these receivers could cover practically 
the whole Antarctic sheet depth for EeV showers. 

\section{Implications for a large radio antenna complex}

A large effective volume for neutrino detection can be obtained with an array of such 
receiver systems. The separation between antennas is a critical parameter for estimating 
the viability of the idea and it is directly related to the angular spread of the pulse at 
the Cherenkov peak \cite{Markov}.
The most important implication of the LPM effect for these showers is 
precisely the reduction in angular width. The pulses will become significantly narrower for 
energies above $\sim 20$~PeV and consequently the antenna separation should become smaller. 
The constraint becomes most severe at the operation frequencies used in \cite{Zas}. A 20 EeV
shower will emit a pulse with an angular half-width reduced by about a factor of 10 with 
respect to PeV showers, i.e. about 2~mrad at 1 GHz. Fortunately the width rises as the 
operating frequency decreases, making lower frequencies more appropriate for detection of 
radio pulses from EeV showers. The significant
reduction in the power received by each antenna when considering low frequencies, 
associated to the frequency dependence of
Cherenkov radiation, is compensated by the fact
that the pulse power scales with the
square of shower energy. Also there is  
an extra benefit in going to lower
operational frequencies because the lower frequency component is less attenuated and 
is less sensitive to temperature \cite{jelley}. 

For an actual design the array parameters such as antenna separation, noise level, 
operating frequency and bandwidth of each antenna are to be optimized for largest 
effective volume and minimal cost. The optimization will depend strongly on the shape 
of the neutrino flux to be measured, but it is clear that EeV neutrino sources can allow 
sparse arrays of antennas in the 100 MHz range and below, which may turn out to be 
cheaper.
Multichannel measurements may be 
advisable for several reasons. Firstly they will reduce background noise (mostly man made) 
and secondly they will allow a better reconstruction of the pulse angular-frequency 
spectrum. The distributions of the electromagnetic pulses in the array can be fitted to 
theoretical predictions, what should allow energy reconstruction in a clear analogy to 
extensive air shower arrays. 

In summary EeV showers produce electric pulses that scale with energy in the Cherenkov direction but have a significantly narrower angular distribution. The technique continues to be promising from the point of view of EeV 
neutrinos if lower frequencies are used, and may allow the detection of the low flux expected from the interactions of cosmic rays with 
the cosmic microwave background. Such a detector could also improve bounds on different 
models for cosmic ray acceleration and open the possibility to further explore the Universe 
in this energy range. It should be stressed however that the technique has not yet been 
proven and that difficulties are anticipated\cite{jelley}. The attractive prospects for neutrino 
detection however justify the experimental effort that is required to test these ideas.

\vskip 0.5 cm
{\bf Acknowledgements:} We thank G.~Parente for suggestions after reading the manuscript and computing assistance and we are grateful to J.J. Blanco-Pillado, 
G.M.~Frichter, F. Halzen, 
A.L. Provorov, J.P. Ralston, T. Stanev and I.M. Zhelehnykh for helpful discussions.
This work was supported in part
by CICYT (AEN96-1773) and by Xunta de Galicia (XUGA-20604A96). One of the authors J. A. thanks the Xunta de Galicia for financial support. 


\newpage

\begin{center}
FIGURE CAPTIONS
\end{center}

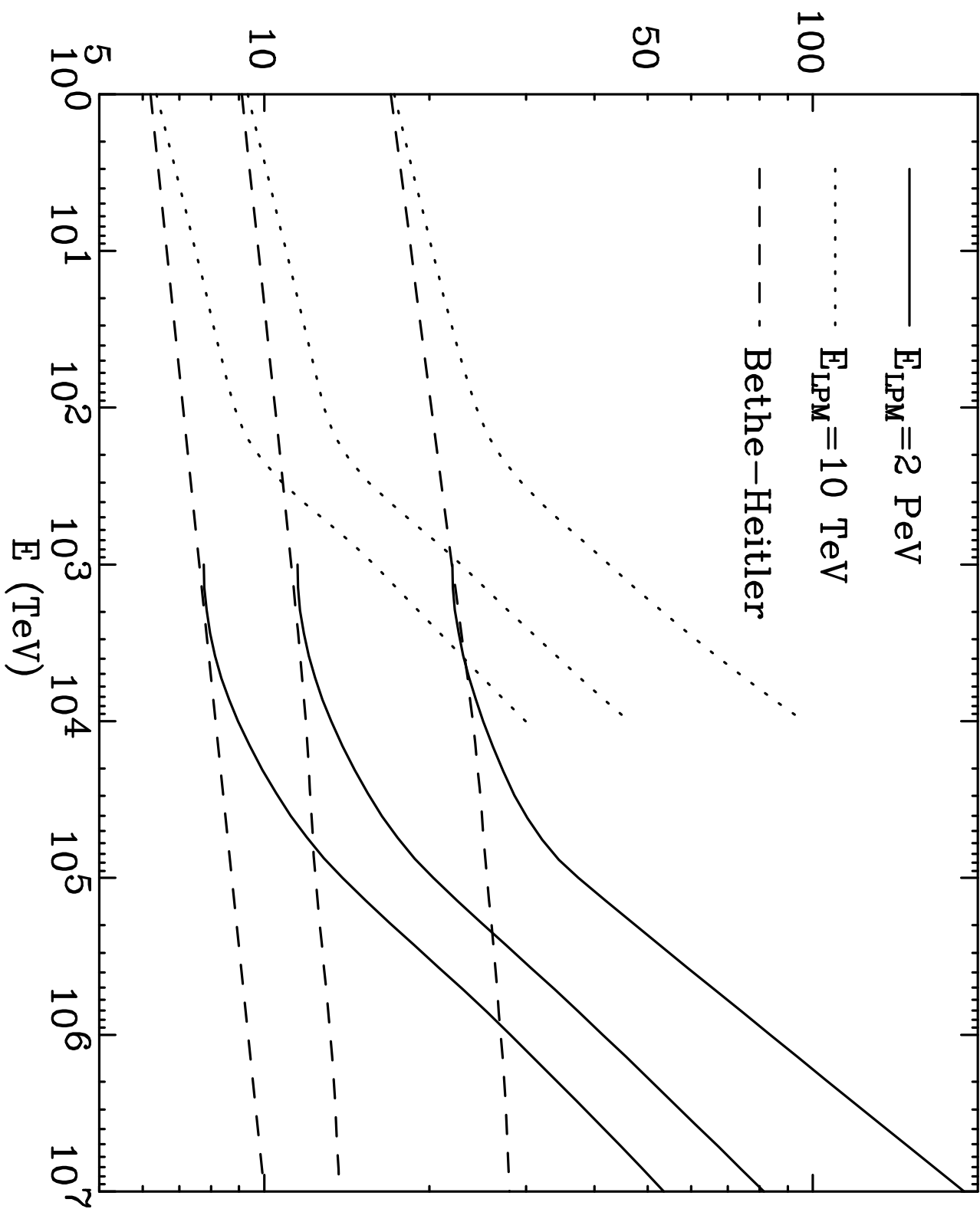
{\bf Figure 1:} Average shower length (defined in text) versus shower energy 
for showers above $E_{LPM}=2$ PeV in ice. The dotted lines represent the effect of 
introducing showers with  
$E_{LPM}=10$ TeV.
From top to bottom curves correspond to shower length defined with $\sim 0.1~N_{max}$, 
$\sim 0.5~N_{max}$ and $\sim 0.7~N_{max}$. 
Dashed lines are corresponding shower 
lengths 
using Greisen's parameterization.

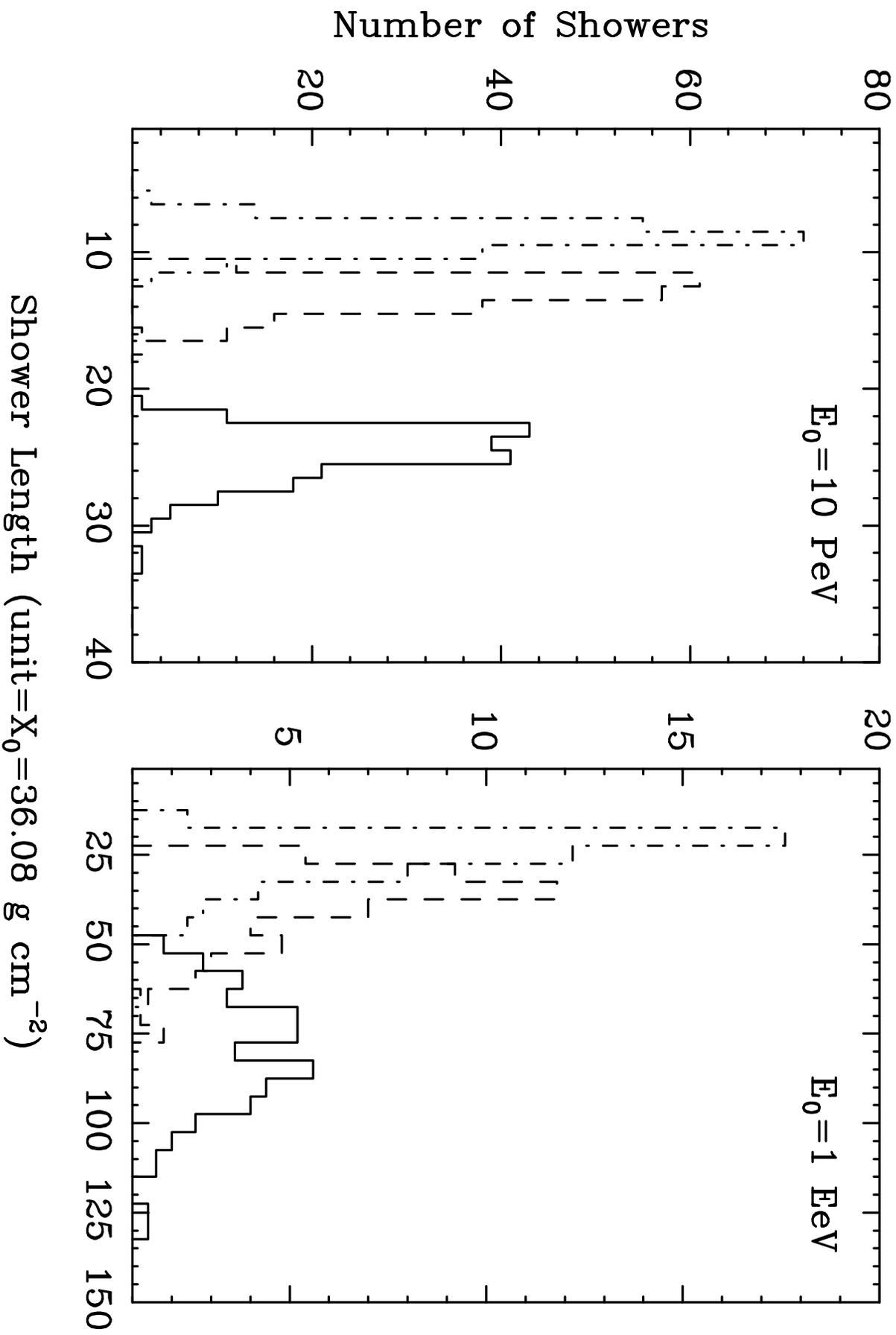
{\bf Figure 2:} Shower length distributions illustrating the increased fluctuations 
with energy using fractions of 0.1 (solid), 0.5 (dashed) and 0.7 (dot-dashed).  

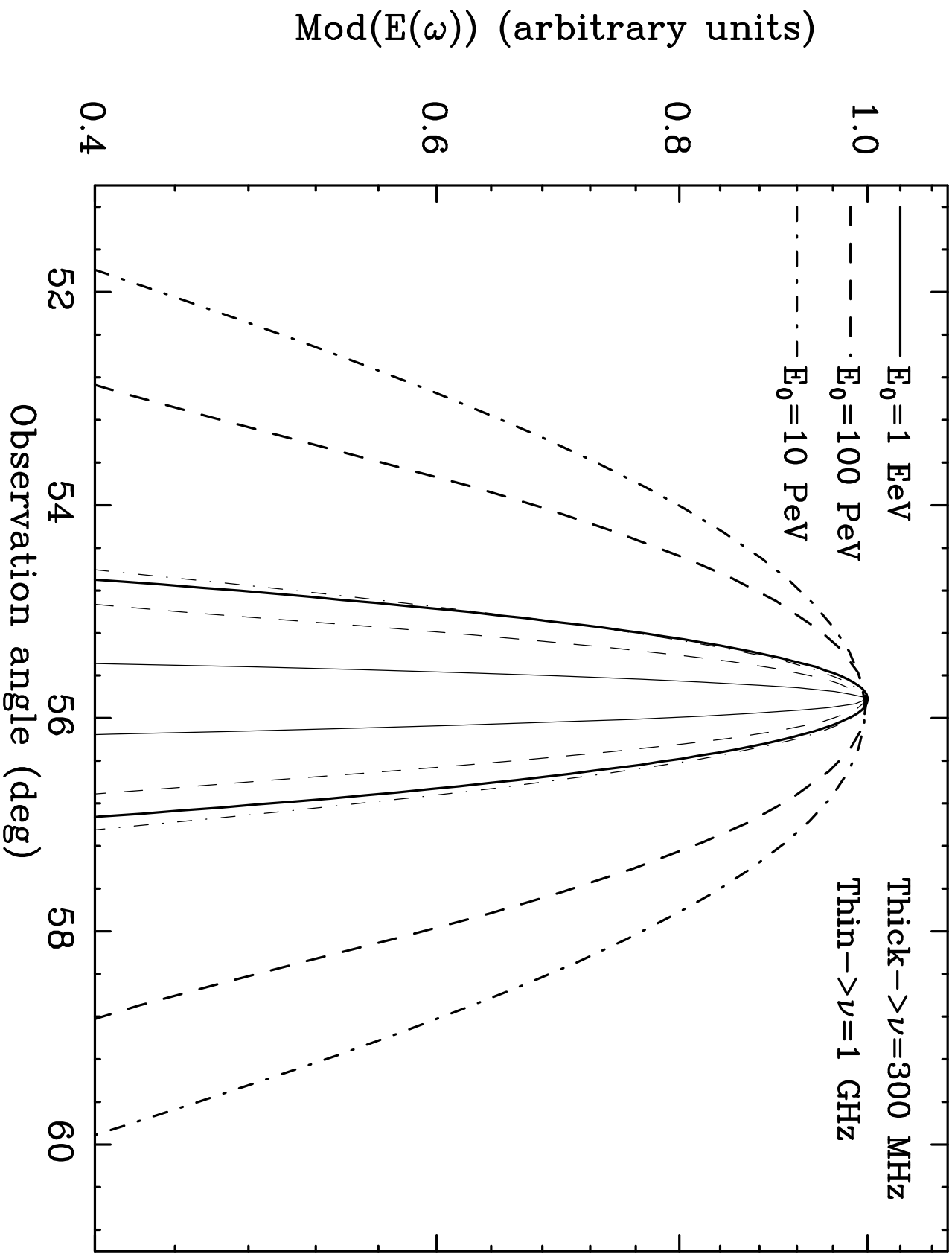
{\bf Figure 3:} Angular distribution of radiopulse around the Cherenkov angle ($56^0$ for 
radiofrequencies) at 300~MHz and 1~GHz, for three different primary energies $E_0$.  

\begin{center}
TABLES
\end{center}

\begin{center}
\begin{tabular}{|c|c|c|c|} \hline
$\nu_{op}$&10 MHz&100 MHz&1 GHz\\ 
$\Delta\nu$&1 MHz&10 MHz&100 MHz\\ \hline\hline
&265 PeV&85 PeV&70 PeV\\ \hline
\end{tabular}
\label{thresh}
\end{center}

{\bf Table I:} Energy thresholds for detection of radiopulses 
for different operation frequencies and bandwidths.


\begin{center}
\begin{tabular}{|c|c|c|c|} \hline
T $\vert$ $\nu$ &10 MHz&100 MHz&1 GHz\\ \hline\hline
$-60^{\rm o}$C&4.4 km&2.6 km&440 m\\ \hline
$-40^{\rm o}$C&660 m&450 m&265 m\\ \hline
$-20^{\rm o}$C&260 m&130 m&70 m\\ \hline
\end{tabular}
\label{table2}
\end{center}

{\bf Table II:} Distance at which power spectrum is halved in ice for different frequencies 
and ice temperatures.


\begin{thebibliography}{999}
%
\bibitem{Auger} For a review see Design Report of the Pierre Auger Collaboration, Fermilab Report, February 1997 and references therein.
%
\bibitem{Biermann} K. Mannheim, {\sl Astropart. Phys.} {\bf 3} (1995) 295; C.T. 
E.~Waxman and J.~N.~Bahcall, Phys. Rev. Lett. {\bf 78} (1997) 2292; 
Hill, D. N. Schramm, and T.P. Walker, {\sl Phys. Rev. } {\bf D36} (1987) 1007.
%
\bibitem{Berezinskii} V.S.~Berezinsky and G.T.~Zatsepin, {\sl Phys.\ Lett.\ 
{\bf B28} (1969) 423}; C.T. Hill and D.N. Schramm, {\sl Phys. Lett.} 
{\bf B131} (1983) 247.
%
\bibitem{zasmoriond} E. Zas in Proc. {\sl Very High Energy Processes in the Universe} 
Moriond 1997.
%
\bibitem{physrep} T.K. Gaisser, F. Halzen, T. Stanev, Phys. Rep. 258 (1995) 
173 and references therein. 
%
\bibitem{Amanda} F. Halzen in Proc. of the {\sl Int. Workshop on 
Aspects of Dark Matter in Astrophysics and Particle Physics}, Heidelberg, 
Germany, September 1996; BAIKAL Collaboration, I. Sokalski and C. Spiering,
The BAIKAL Neutrino Telescope NT-200, BAIKAL 92-03 (1992).
%
\bibitem{zele} I.M. Zeleznykh, {\sl Proc.\ XXIth International
Cosmic Ray Conference} (Adelaide, 1989), Vol.~6, p.~528--533.
%
\bibitem{aska} G.A. Askar'yan, Soviet Physics JETP {\bf 14,2} (1962)
441;  48 (1965) 988. 
%
\bibitem{price} P.B. Price, Astroparticle Physics 5 (1996) 43.
%
\bibitem{Zas} F.Halzen, E.Zas, T.Stanev, Phys. Lett. B 257 (1991) 432; 
E.Zas, F.Halzen, T.Stanev, Phys. Rev. D 45 (1992) 362. 
%
\bibitem{jelley} J.V. Jelley, Astroparticle Physics 5 (1996) 255 and references therein.
%
\bibitem{RICE} G.M. Frichter {\it private communication}.
%
\bibitem{jaime} R.D. Dagkesamansky and I.M. Zheleznykh, Radioastronomical Method of
the Neutrino and Hadron Detection, in Proc. of the ICRR 
International Symposium: Astrophysical Aspects of the most energetic 
Cosmic Rays (Kofu, Japan, November 1990), 
eds. M. Nagano and F. Takahara (World Scientific, 1991) p.373;
 J. Alvarez-Mu\~niz and E. Zas, in Proc. of Workshop 
in High Energy Neutrino Astrophysics (HENA 96), Ed. T.K.~Weiler 1996.
%
\bibitem{reno} C. Quigg, M.H. Reno, T.P. Walker, 
Phys. Rev. Lett. 57 (1986) 774.
%
\bibitem{LPM} L. Landau and I. Pomeranchuk, {\sl Dokl.\ Akad.\
Nauk\ SSSR}
{\bf 92} (1953) 535; {\bf 92} (1935) 735.
%
\bibitem{migd} A.B. Migdal, Phys. Rev.
103 (1956) 1811;  Zh. Eksp. Teor. Fiz.
32 (1957) 633 [  Sov. Phys. JETP
5 (1957) 527].
%
\bibitem{konishi} E. Konishi, A. Adachi, N. Takahashi and A. Misaki, 
J. Phys. G: Nucl. Part. 17 (1991) 719.
%
\bibitem{stanvan} T. Stanev, Ch Vankov, R.E. Streitmatter,
R.W. Ellsworth and T. Bowen, Phys. Rev. D 25 (1982) 1291.
%
\bibitem{Misaki} A. Misaki, Phys. Rev. D 40 (1989) 3086, Nucl. Phys. B (Proc. Suppl.) 33 (1993) 192.
%
\bibitem{Nishimura} J. Nishimura, {\sl Theory of Cascade Showers}, (ed.\
Flugge,~S.) Springer, Berlin (1967) {\sl Handbuch der Physik} Bd.\
{\bf XLVI}/2.
%
\bibitem{NKG} K. Greisen, in: Prog. of Cosmic Ray Phys., ed.
J.G. Wilson, Vol.~III, (North Holland Publ. Co., Amsterdam, 1956) p.1., K.~Kamata and J.~Nishimura, Prog.~Theor.~Phys. (Kyoto) Suppl. {\bf 6} (1958) 93.
%
\bibitem{ralpm} J.P. Ralston and D.W. McKay, Comparing Coherent Microwave Emission from LPM 
and BH
showers, in: Proc. of High Energy Gamma-Ray Astronomy Conference (Ann Arbor, Mi 1990), ed. James
Matthews
(AIP Conf. Proc. 220) p.295.
%
\bibitem{prov} A.L. Provorov, I.M. Zheleznykh, Astroparticle Physics 4 
(1995) 55.
%
\bibitem{Ralston} G.M. Frichter, J.P. Ralston, D.W. Mc Kay, Phys. Rev. D 53
(1996) 1684.
%
\bibitem{bold} Boldyrev,~I.N., Gusev,~G.A., Markov,~M.A., Provorov,~A.L. 
and Zeleznykh,~I.M., {\sl Proc.\ XXth International Cosmic Ray Conference}
(Moscow, 1987), Vol.~6, p.~472.      
%
\bibitem{Icebook} P. Hobbs, Ice Physics, Clarendon Press, Oxford, (1974). 
%
\bibitem{Markov} M.A. Markov, I.M. Zheleznykh, Nucl. Instr. and Methods in Phys. Research 
A248 (1986) 242.
%
\end{thebibliography}
\end{document}